\documentclass[
]{ceurart}

\usepackage{listings}
\begin{document}

\copyrightyear{2024}
\copyrightclause{Copyright for this paper by its authors.
  Use permitted under Creative Commons License Attribution 4.0
  International (CC BY 4.0).}

\title{Advancing GenAI Assisted Programming--- \\A Comparative Study on Prompt Efficiency and \\
Code Quality Between GPT-4 and GLM-4}


\author[1,2]{Angus YANG}[
email=ayan5942@uni.sydney.edu.au,
]
\fnmark[1]
\address[1]{School of Computer Science, Faculty of Engineering, The University of Sydney}
\address[2]{DeepBlue Technology (Shanghai)}

\author[2,3]{Zehan LI}[
email=18612508125@163.com,
]
\cormark[1]
\address[3]{AI Committee, China Creative Studies Institute}

\author[3,4]{Jie LI}[
email=lijiecs@sjtu.edu.cn,
]
\address[4]{Department of Computer Science and Engineering, Shanghai Jiao Tong University}

\cortext[1]{To whom correspondence should be addressed; Email: 18612508125@163.com }
\fntext[1]{Work done while interning at DeepBlue Technology (Shanghai Headquarters)}

\begin{abstract}
This study aims to explore the best practices for utilizing GenAI as a programming tool, through a comparative analysis between GPT-4 and GLM-4. By evaluating prompting strategies at different levels of complexity, we identify that simplest and straightforward prompting strategy yields best code generation results. Additionally, adding a CoT-like preliminary confirmation step would further increase the success rate. Our results reveal that while GPT-4 marginally outperforms GLM-4, the difference is minimal for average users. In our simplified evaluation model, we see a remarkable 30 to 100-fold increase in code generation efficiency over traditional coding norms. Our GenAI Coding Workshop highlights the effectiveness and accessibility of the prompting methodology developed in this study. We observe that GenAI-assisted coding would trigger a paradigm shift in programming landscape, which necessitates developers to take on new roles revolving around supervising and guiding GenAI, and to focus more on setting high-level objectives and engaging more towards innovation.
\end{abstract}


\maketitle

\section{Introduction}

The advent of Generative Artificial Intelligence (GenAI) has marked a pivotal moment in the evolution of software development, ushering in an era of unprecedented potential for automation and innovation. Since the release of ChatGPT 3.5 in November 2022 and GPT-4 in March 2023, the landscape of GenAI has witnessed rapid advancements, both in terms of the foundational model capabilities and their applications across diverse fields (OpenAI, 2023)\cite{openai2023gpt4}. These advancements have significantly expanded the horizons of what is achievable with AI, showcasing the ability to enhance efficiency and elevate the quality of outcomes across various domains (T. Eloundou et al., 2023)\cite{eloundou2023gpts}. Among the myriad applications of GenAI, its integration into programming practices stands out as one of the most transformative, promising to redefine the paradigms of software development (Jaber et. al., 2023; Peng et al. 2023)\cite{beganovic2023methods}\cite{peng2023impact}.

The application of GenAI in programming has exemplified impressive potential, demonstrating that when leveraged effectively, it can significantly augment the coding process, making it more efficient and intuitive. The capacity of GenAI to understand complex instructions, generate code, and assist in debugging has made it an invaluable tool for developers, enabling them to focus on higher-level design and problem-solving aspects of software development. However, the integration of GenAI into programming is not without its challenges (Hendler, 2023)\cite{hendler2023understandlimits}. Key among these are determining the optimal ways to prompt the AI for the best quality of coding, evaluating the quality of the code generated, and establishing effective collaboration practices between human developers and AI systems. Additionally, selecting the appropriate AI platforms that align with specific project needs poses a significant decision-making challenge (Yu et al., 2024)\cite{yu2024codereval}.

Recognizing these challenges, our research aims to design a generalizable and accessible methodology to evaluate the efficiency and quality of GenAI-assisted coding. Through conducting a comparative study across different major GenAI platforms, we seek to derive a set of general guidelines or best practices for leveraging GenAI in programming, of which the importance cannot be overstated. As GenAI continues to evolve, setting operational norms and guidelines becomes crucial to maximizing its benefits while mitigating potential drawbacks (Hamza et al. 2023)\cite{hamza2023human}. In doing so, this study contributes to the broader discourse on the future of software development, highlighting the transformative impact of GenAI and setting the stage for a new era of programming practices.

\section{Literature Review}
\subsection{Overview of Existing Studies on GenAI Applications in Programming}
Generative Artificial Intelligence (GenAI) has significantly influenced the field of software development, offering tools for code generation, debugging, and even writing documentation. Studies such as those by Weisz et al. (2022) \cite{Weisz_2022} have evaluated AI-supported code translation, demonstrating how GenAI can aid developers in producing code with fewer errors and in a more efficient manner. Similarly, Peng et al. (2023) \cite{peng2023impact}have shown that tools like GitHub Copilot can improve developer productivity by assisting in rapid code generation, indicating a positive impact on the speed and efficiency of software development tasks. Hamza et al. (2023)\cite{hamza2023human} investigated the dynamics of human-AI collaboration in software engineering, focusing on how leveraging ChatGPT can enhance coding efficiency and optimization. Their research identified challenges and proposed mitigation methods for integrating AI into development processes, including addressing security risks and enhancing human oversight. 

\subsection{Analysis of the Impact of Prompt Scheme, Complexity, and Clarity on GenAI Performance}
The effectiveness of GenAI in programming is highly dependent on the prompt's scheme, complexity, and clarity. Jonsson and Tholander (2022)\cite{jonsson2022cracking} explored the use of GenAI for educational purposes, highlighting how the design of prompts affects the tool's ability to support creative programming tasks. White et al. (2023)\cite{white2023chatgpt} investigated prompt pattern design techniques with LLMs in software engineering, and proposed effective prompt patterns that would secure a more rapid and higher quality code generation. This underscores the importance of prompt engineering in maximizing the potential of GenAI tools for coding, where clear and well-structured prompts can significantly enhance the quality and relevance of generated code (Hamza et al., 2023)\cite{hamza2023human}. 

\subsection{Comparative Analysis of Different LLMs in Programming Contexts}
Yu et al. (2024)\cite{yu2024codereval} recently proposed an effective benchmark (CoderEval) to evaluate pragmatic code generation scenarios, open source or proprietary ones, for three SOTA code generation models (Code-Gen, PanGu-Coder, and ChatGPT). Nevertheless, focusing on GPT-4 and GLM-4 offers a compelling narrative within this broader context. OpenAI's GPT-4, as one of the most advanced and most accessible foundation models to date, serves as a benchmark in many comparative studies. Its capabilities not only encompass code generation but also extend to debugging, providing explanations, and even writing tests, setting a high standard for what is achievable with current AI technologies (OpenAI, 2023)\cite{openai2023gpt4}.  On the other hand, GLM-4 emerges as one of the most capable Chinese foundation models, recently introduced in January of 2024 (ZHIPU-AI, 2024)\cite{ZHIPU24DevDay}. GLM-4's training on diverse datasets, including those in Chinese, makes it particularly adept at understanding and generating code in contexts where cultural and linguistic nuances play a crucial role (Zeng et al., 2022; ZHIPU-AI, 2024)\cite{zeng2023glm130b}\cite{ZHIPU24DevDay}. This makes the comparison between GPT-4 and GLM-4 not only interesting but also highly relevant, especially for users and developers operating within the Chinese tech ecosystem.  While direct comparative studies between GPT-4 and GLM-4 in programming contexts are scarce, which makes this study more valuable and necessary.

\subsection{Gap in Literature Regarding Operational Norms for GenAI-Assisted Programming}
One notable gap in the literature is the lack of established operational norms for GenAI-assisted programming. While studies have begun to explore the efficacy and applications of GenAI tools in software development, there is a need for comprehensive guidelines and best practices that address how to integrate these tools effectively into programming workflows. This includes considerations for prompt design, model selection, and the evaluation of generated code quality, as well as ethical considerations related to code originality and security.

\subsection{Concluding Remarks}
The literature on GenAI applications in programming highlights the transformative potential of these tools in enhancing developer productivity and creativity. However, the effectiveness of GenAI is contingent upon the quality of prompts and the characteristics of the underlying models. There remains a significant need for research that establishes operational norms and best practices for integrating GenAI tools into software development processes, ensuring their responsible and effective use.

\section{Methods}
\subsection{Research Design}
This study aims to explore the best practices for utilizing GenAI as a programming tool. Our primary goal is to establish operational norms for GenAI-assisted programming via a comparative analysis. Specifically, we would delve into effective prompt strategies, evaluation methods for GenAI platforms, approaches for assessing AI-generated code, and evaluating coding efficiency improvement. The key variables under examination are: 1) complexity and clarity of prompt wording, and 2) comparative performance of different LLMs (GPT-4 versus GLM-4).

Our research will focus on using GenAI to generate code for program modules, evaluating various prompt strategies and the code generation capabilities of the underlying large language models (LLMs). For this purpose, we have selected the classic arcade game "Snake" as our subject module, based on several considerations. Firstly, the simplicity of "Snake" makes it an exemplary case for studying algorithm optimization and efficiency, thereby allowing for a concrete assessment of code effectiveness, as highlighted by Yeh et al. (2016)\cite{yeh2016snake}. The iterative development process inherent in games like "Snake", which necessitates numerous enhancements and refinements, underscores the significance of code readability and maintainability—a factor critical to the quality of game code as emphasized by Tashtoush et al. (2013)\cite{tashtoush2013business}. Readability and maintainability are essential for facilitating updates and improvements to the game, thus establishing "Snake" as an invaluable model for examining these elements in code development. Moreover, employing arcade games as educational tools not only boosts students' motivation but also aids in comprehending complex programming concepts and encourages active participation in programming tasks, as discussed by Theodoraki and Xinogalos (2014)\cite{theodoraki2014studying}. The development skills acquired through "Snake" are highly adaptable to more sophisticated and practical applications, including advanced software development and game design, further affirming the game's applicability and significance beyond its primary context, as reiterated by Yeh et al. (2016)\cite{yeh2016snake}.
\begin{figure}[h!]
    \centering
    \includegraphics[width=1\linewidth]{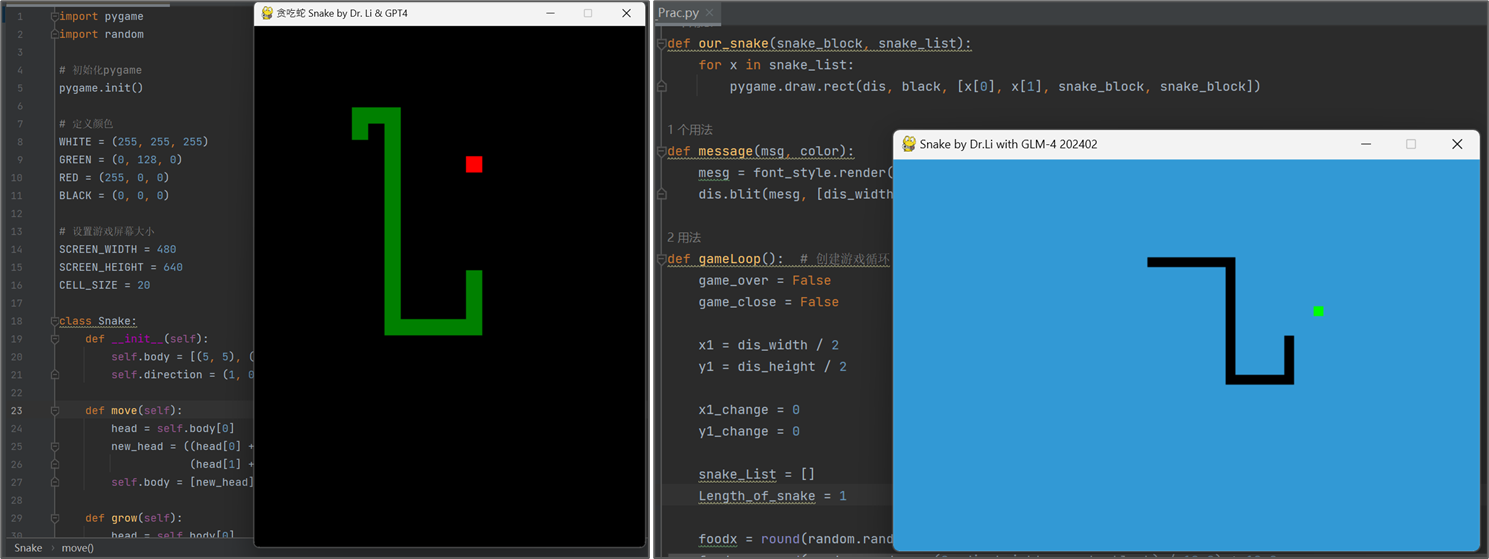}
    \caption{Representative screenshots of Snake games programmed in Python v3.11, assisted by GenAI: GPT-4 (left) and GLM-4 (right)}
    \label{Figure 1}
\end{figure}

\subsection{Evaluation Criteria}
The evaluation criteria for this study are prioritized as follows:

\textbf{1) Success Rate:} The primary criterion is the success rate of one-shot prompts in generating a robust core module for the Snake program across various prompt strategies. This metric assesses the effectiveness of different prompt approaches in producing a functional program on the first attempt.

\textbf{2) Debugging Efficiency:} In cases where one-shot prompt generation does not result in a viable core module, the efficiency of GenAI-assisted debugging necessary for successful program execution becomes the secondary criterion. This evaluates the supportiveness of GenAI in streamlining the debugging process.

\textbf{3) Code Conciseness and Readability: }Once the previous criteria are met equivalently, the focus shifts to the conciseness and readability of the generated code. This criterion underscores the importance of generating code that is not only functional but also clean and easy to understand.

\textbf{4) Functionality Completeness and Richness:} Lastly, if all preceding criteria are satisfied comparably, the comparison of the generated program's completeness and the richness of its functionality takes precedence. This assesses the extent to which the generated code encompasses a comprehensive set of features and capabilities.

These criteria collectively aim to evaluate the generated code not just on its immediate functionality, but also on its efficiency, maintainability, and the breadth of its capabilities, ensuring a holistic assessment of GenAI's contribution to programming tasks.

\subsection{Prompt Levels Design}
\subsubsection{One-shot prompt}
To evaluate ChatGPT's capability in code generation relative to prompt complexity, we have designed four distinct prompts that embody increasing levels of complexity. These prompts are intended to examine how varying approaches to prompting interact with SOTA LLMs in terms of code generation capabilities. The prompts are structured as follows for one-shot prompt:

\textbf{First Level:} This prompt offers a vague description of a straightforward task, setting the foundation for understanding the model's response to minimal input detail.

\textbf{Prompt 1:} Generate code for a snake game in python.

\textbf{Second Level:} Building on the first, this prompt includes a simple task description but with slightly more detail, providing a clearer direction based on the initial vague description.

\textbf{Prompt 2:} Generate code for a snake game in python, where the player controls an unstoppable snake, the snake grow longer when it eats the dots on the screen. The player loses the game when the snake’s head comes in contact with the edge of the screen or the snake’s body.

\textbf{Third Level:} The third prompt introduces further specifications and constraints, aiming to assess the model's ability to adhere to more complex requirements.

\textbf{Prompt 3:} Generate code for a snake game in python, where the player controls an unstoppable snake, the snake grow longer when it eats the dots on the screen. The player loses the game when the snake’s head comes in contact with the edge of the screen or the snake’s body. Make sure the code follows SOLID principles, include a page that provide the option to restart when player loses the game.

\textbf{Forth Level:} This prompt simplifies the description by integrating essential sections of pseudo code, focusing on the critical components needed for the task.

\textbf{Prompt 4:} Generate code for a snake game in python, where the player controls an unstoppable snake, the snake grow longer when it eats the dots on the screen. The player loses the game when the snake’s head comes in contact with the edge of the screen or the snake’s body. For player to control the snake, include function or functions to read player input and change the direction of the snake for the corresponding input. Include the function to get generate the dot onto a place that’s not a part of the snake. Include a function to extend the length of the snake once it touches the dot, the dot should then be removed and a new dot should be placed. Include a function to check the collision between snake’s head and its body and wall, stop the game if collision is confirmed.

This structured approach allows us to observe the variation in AI-generated code quality in response to different prompt complexities. We hypothesize that prompts with greater detail and specificity will yield higher-quality outputs, demonstrating the importance of prompt design in leveraging LLMs for code generation.

\subsubsection{Follow-up prompt setup}
Evaluating the proficiency of LLMs in coding tasks necessitates an examination of their ability to modify and refactor code in response to feedback on the success or failure of their initial outputs. This method assesses not only the LLM's initial capability to generate code but also its aptitude for iterative enhancement and adjustment based on evaluative input. When provided with feedback on the performance of its previously generated code, the LLM is tasked with comprehending and applying this information to execute informed revisions. The structure for follow-up prompts is as follows:

\textbf{For Success:} “The program worked successfully. Attempt generating code with alternative approaches or solutions.”

\textbf{For Failure with an Error Message:} “Received error message: [insert error message here]. Please make adjustments and try again.”

\textbf{For Failure without an Error Message:} “The program failed to execute as expected. Attempt to identify and correct the issue before trying again.”

This setup is designed to simulate a realistic coding workflow, where developers often iterate on their code based on the results of testing and debugging. By adopting this approach, we aim to closely examine the LLM's capacity for engaging in a developmental feedback loop, highlighting its potential for learning and adaptation in programming tasks.

\subsection{Data Collection}
To ensure a robust statistical analysis, each of the four different prompts will be tested 20 times independently, resulting in a substantial sample size conducive to reliable statistical inferences. For every prompt test, two follow-up prompts will be administered, yielding three distinct outcomes per test. Consequently, this testing protocol will generate a total of 240 outcomes.

For the purpose of data collection, these 240 outcomes will be categorized into one of the following four types:

\textbf{Pass:} indicates the program ran successfully with all the rules of the Snake game correctly implemented, demonstrating a complete and functional game.

\textbf{Failure Type 1:} GenAI's output is incomplete, such as instances where the code contains unfilled blanks requiring user intervention, the code is not delivered in a single contiguous block (e.g., fragmented across multiple code boxes), or no code is generated at all.

\textbf{Failure Type 2:} GenAI produces code that appears complete, but execution of the code fails to launch a game window due to programming errors, indicating issues with the code's functionality.

\textbf{Failure Type 3:} a game window is successfully launched, but interaction within the game does not adhere to the established rules of the Snake game (e.g., the game does not terminate when the snake collides with a wall), pointing to deficiencies in game logic or rule implementation.

\subsection{Experiment Procedure}
The experiment procedure is outlined in a step-by-step format to ensure clarity and consistency in the execution of tests:

\textbf{Step 1. Initiate a New Session:} Begin a new conversation with the LLM.

\textbf{Step 2. Input a Prompt:} Enter one of the predetermined prompts into the LLM.

\textbf{Step 3. Execute Generated Code:} Run the AI-generated code in a Python environment.

\textbf{Step 4. Record Outcome:} Document the results of the code execution, noting whether it falls into the Pass category or one of the Failure types.

\textbf{Step 5. Perform Follow-up Prompts:}

   a. Based on the initial outcome, input the corresponding follow-up prompt into the LLM.
   
   b. Copy the newly generated code and run the program in a Python environment.
   
   c. Record the outcome of this execution.
   
   d. Repeat steps 5a to 5c for a second iteration of follow-up prompts.

\textbf{Step 6. Repeat Testing Cycle:} Continue the process from steps 1 to 5 for a total of 20 iterations for each prompt.

\textbf{Test All Four Levels of Prompts:} follow step 1-6 above with all the four levels prompts, will yield a total of 240 (60 outcomes x 4 levels) records, for each LLM.

\section{Results}
\subsection{Results for GPT-4}

   \begin{table}[h!]
       \centering
\caption{Code Generation Success Rates for GPT-4}
\label{Table 1}
       \begin{tabular}{cccccccccclll} 
       \toprule
            Prompt&  p1s1&  p1s2&  p1s3&  p2s1&  p2s2&  p2s3&  p3s1&  p3s2&  p3s3& p4s1& p4s2&p4s3\\ 
        \midrule
            GPT-4&  90\%&  50\%&  55\%&  70\%&  35\%&  70\%&  45\%&  30\%&  20\%& 45\%& 35\%&45\%\\ 
        \bottomrule
       \end{tabular}

   \end{table}

In Table 1, "p" represents the prompt level, and "s" denotes the sequence of interaction. Thus, "p1" to "p4" correspond to the first through fourth levels of prompt complexity as described in Section 3.3.1. The sequence following "p", denoted by "s1" to "s3", refers to the interaction sequence, where "s1" is the initial input, "s2" is the first follow-up prompt, and "s3" is the second follow-up prompt.
\begin{figure}
    \centering
    \includegraphics[width=1\linewidth]{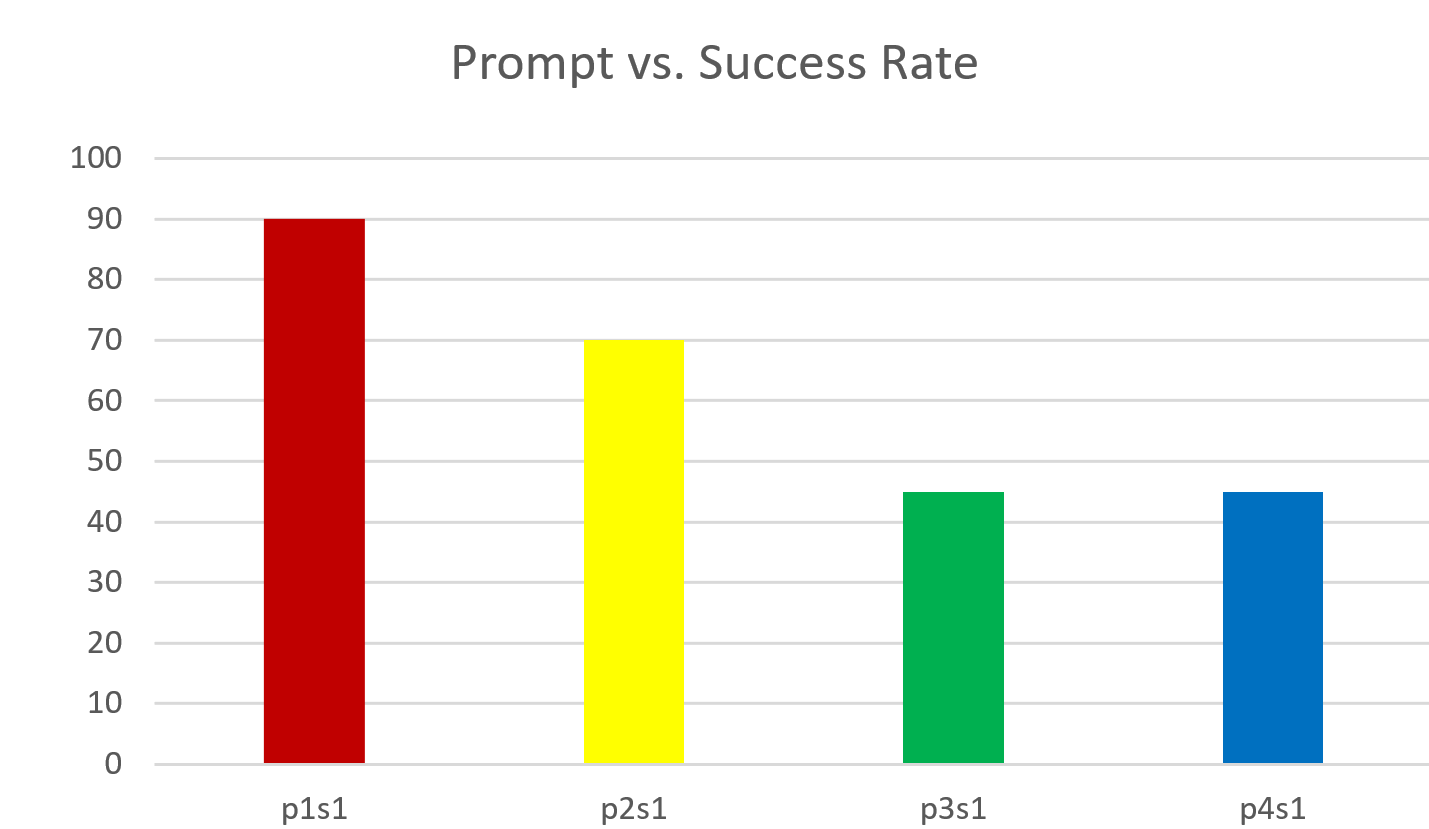}
    \caption{One-shot prompt success rate for GPT-4}
    \label{Figure 2}
\end{figure}

Figure 2 clearly illustrates that the simplest prompt, "Prompt 1", yielded the highest reliability in code generation, achieving a success rate of 90\%. This contrasts markedly with the initial attempt at "Prompt 4", which registered a much lower success rate of 45\%. As the prompts increase in complexity, the likelihood of the generated code meeting the set requirements declines, with "Prompt 1" averaging a success rate of 65\% compared to Prompt 4's approximate rate of 42\%. A plausible explanation for this trend is that simpler prompts enable more efficient processing. They utilize fewer tokens, which in turn allows the model to more effectively dedicate its computational power to producing accurate and relevant code.

\begin{table}[h!]
\centering
\caption{Failure type distribution}
\label{Table 2}
\begin{tabular}{l rrrrrrrrrrrr}
\hline
Type & p1s1 & p1s2 & p1s3 & p2s1 & p2s2 & p2s3 & p3s1 & p3s2 & p3s3 & p4s1 & p4s2 & p4s3 \\
\hline
Failure1 & 100\% & 40\% & 22\% & 50\% & 46\% & 67\% & 27\% & 43\% & 25\% & 0 & 23\% & 9\% \\
Failure2 & 0 & 60\% & 33\% & 17\% & 31\% & 33\% & 18\% & 7\% & 38\% & 64\% & 23\% & 64\% \\
Failure3 & 0 & 0 & 44\% & 33\% & 23\% & 0 & 55\% & 50\% & 38\% & 36\% & 54\% & 27\% \\
\hline

\end{tabular}

\end{table}

As shown in Table 2, a notable concentration of Failure Type 1 was observed during the tests of prompts p2s1-p2s3, occurring consecutively within the 24 hours allocated for the p2 series tests. We noticed LLM outputs featuring multiple code boxes and some lines of code appearing outside of these boxes. This phenomenon, not observed in subsequent tests, may have been an outlier resulting from temporary adjustments to the GPT-4 model by OpenAI.

Moreover, there is a discernible trend where Failure Type 1 decreases as the complexity of prompts increases, alongside a subtle rise in Failure Type 3 instances. This suggests that more complex prompts may indeed prompt the LLM to more frequently produce complete code outputs, albeit with an increased likelihood of encountering Type 3 failures. This trend implies that while detailed prompts enhance the model's ability to generate complete outputs, they also elevate the risk of generating code that, while complete, does not function correctly within the intended game rules.

\subsection{Results for GLM-4}

\begin{table}[h!]
\centering
\caption{Code Generation Success Rates for GLM-4}
\label{Table 3}
\begin{tabular}{l l l l l l l l l l l l l}
\hline
Prompt & p1s1 & p1s2 & p1s3 & p2s1 & p2s2 & p2s3 & p3s1 & p3s2 & p3s3 & p4s1 & p4s2 & p4s3 \\
\hline
GLM-4 & 90\% & 10\% & 20\% & 30\% & 30\% & 35\% & 35\% & 20\% & 35\% & 45\% & 35\% & 35\% \\
\hline

\end{tabular}

\end{table}
\begin{figure}
    \centering
    \includegraphics[width=1\linewidth]{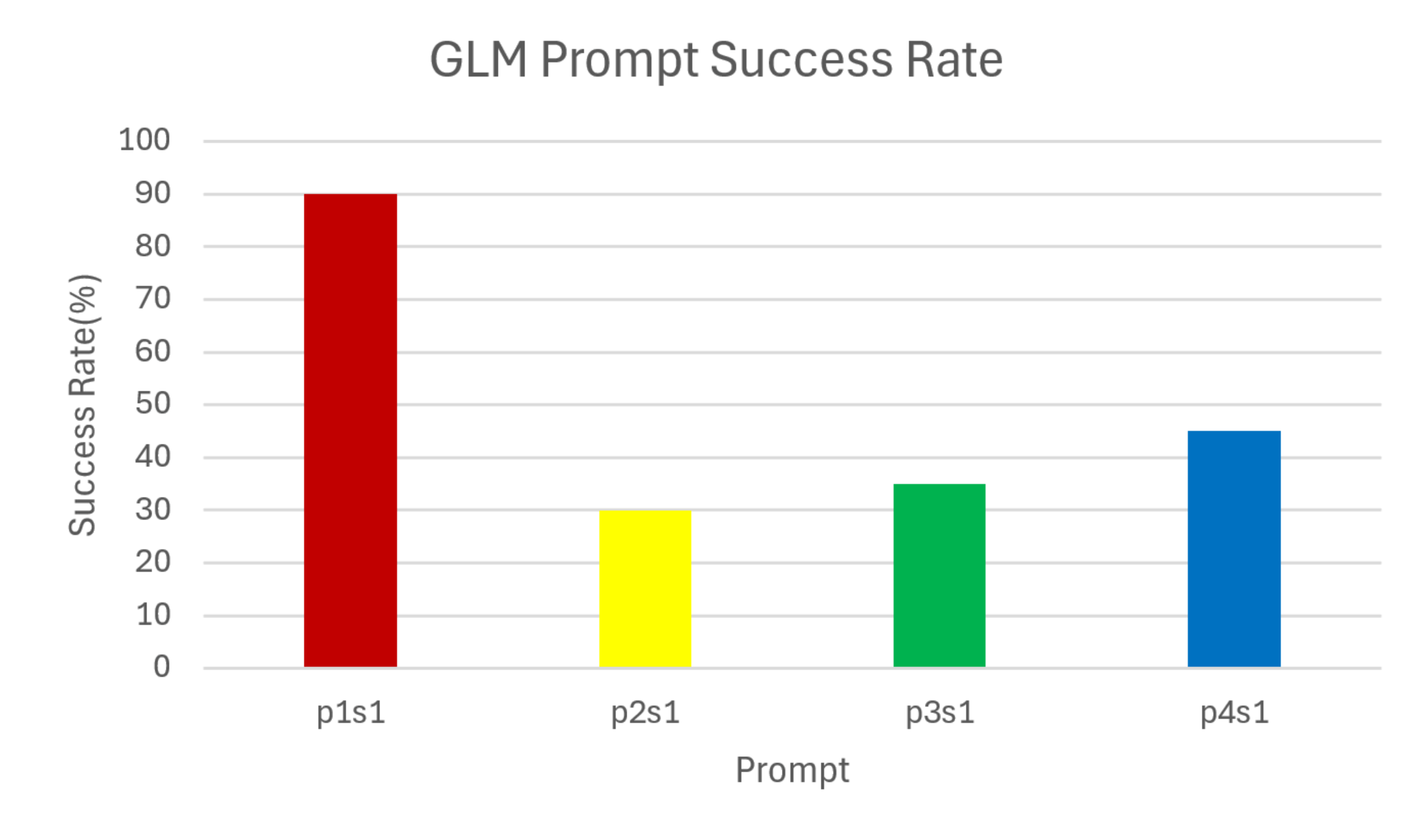}
    \caption{One-shot prompt success rate for GLM-4}
    \label{Figure 3}
\end{figure}

From Table 3 and Figure 3 above, we can see that GLM-4's performance trends show similarities to those observed with GPT-4, displaying an overall decline in success rates across the prompts. Initially, the first attempt at Prompt 1 boasts a high success rate of 90\%, which significantly drops to 45\% by the first attempt at the final prompt. 

A notable issue with GLM-4 is the frequent occurrence of a specific error where the generated code is prematurely truncated at the final lines. This error, occurring as GLM-4 attempts to generate complete code and mistakenly believes it has succeeded without issuing an error message or prompting for regeneration, is classified as a Failure Type 2, representing "Failed content of code generation."

Moreover, a significant number of Failure Type 1 instances were observed in the follow-up prompts for Prompt 1. This likely highlights GLM-4's difficulty in interpreting vague terms, as the follow-up prompts related to the initially vague Prompt 1 are also relatively indistinct. This suggests a potential weakness in GLM-4's ability to process and respond to less specific prompts effectively.

\subsection{Code Generation Performance: GPT-4 vs GLM-4}
In this comparative analysis of code generation capabilities, GPT-4 slightly outperforms GLM-4, showcasing a higher overall success rate across a range of code generation tasks (Note: full results and representative GenAI-generated code samples please refer to Appendix 1.). Notably, GLM-4 reaches a success rate comparable to GPT-4 with the final and presumably most complex prompt, despite GLM-4's consistent struggle with Type 2 failures, where the generated code tends to be incomplete. Our observations indicate that GLM-4's successes are predominantly achieved through the exclusive use of the Pygame library, whereas GPT-4 diversifies its approach, experimenting with Turtle and other libraries, particularly in prompts 3 and 4. These prompts yield lower success rates compared to those utilizing the Pygame library. This suggests that while GPT-4 currently holds the advantage in overall performance, GLM-4's specialized use of Pygame highlights a potential area of strength. Should GLM-4 address and mitigate its frequent Type 2 failures, there is a possibility for it to narrow the performance gap with GPT-4 or potentially exceed GPT-4's performance in specific areas.

\subsection{Enhancing One-shot Success Rate with Preliminary Confirmation}
Our study identifies a significant increase in one-shot success rates for LLMs following a preliminary "confirmation round". By posing a basic question, such as "Do you know the classic arcade game Snake?" prior to code generation, we effectively prime the LLM for the task at hand. This method improves the LLM's understanding of the task, leading to a higher success rate and coding efficiency.

This enhancement can be linked to the Chain-of-Thought (CoT) mechanism, as illustrated by Wei et al. (2022)\cite{wei2023chainofthought}. CoT prompting has been shown to boost LLMs' performance across various reasoning tasks significantly. By initiating a CoT process with a contextual question, we set a focused reasoning pathway, encouraging the LLM to generate more accurate outputs through a series of logical steps. This strategy not only aligns the model's focus with specific task objectives but also clarifies the task requirements, reducing ambiguity and optimizing the model's computational efficiency for solving the given problem.

\section{Discussion}
\subsection{Limitations and Generalization of Research Findings}
\textbf{1. Target audience limitation}: The prompting techniques and frameworks developed in this research are primarily tailored for programming beginners and independent developers, rather than for software engineers in well-established IT corporations. The latter group typically adheres to stringent programming standards and possesses vast code repositories, diminishing the likelihood of their reliance on GenAI for coding tasks. Additionally, concerns regarding information security and intellectual property protection present significant barriers for large corporations considering the adoption of GenAI-assisted coding practice at this early stage of GenAI applications.

\textbf{2. Programming language focus:} The scope of our analysis is centered on Python, which inherently restricts our understanding of the models' performance when applied to other programming languages, each characterized by its distinct syntax, idioms, and usage patterns. This means that our conclusions may not fully translate to or adequately represent the capabilities of GenAI in handling the nuances of other languages.

\textbf{3. Prompt selection bias:} The selection of test prompts in our study may exhibit a bias towards certain types of tasks while potentially neglecting edge cases. This could skew the perceived effectiveness of GenAI in varied coding scenarios. Furthermore, the objective assessment of code quality and functionality is challenging without detailed human review, which is necessary to capture the nuances of code performance and adherence to best practices. 

\textbf{4. Technological evolution: }The rapid advancement in AI technology means models such as GPT-4 and GLM-4 are frequently updated, which can change their code generation capabilities. Consequently, the relevance of our conclusions could diminish for future versions of these models. Additionally, with the introduction of new and more capable foundation models, our insights could necessitate reevaluation or may become outdated.  Readers are advised to exercise discretion, considering the dynamic nature of AI technology.

\textbf{5. Generalization of evaluation methodology: }Our evaluation methodology, which employs four prompts of varying complexity to assess LLMs' Python programming capabilities, introduces a nuanced approach and structured framework to the evaluate the capability of AI code generation. The flexibility of this approach suggests it could be adapted for evaluation of other programming languages and across more diversed GenAI platforms.  There's also a conceivable extension of this methodology to evaluate the efficacy of AI-supported code review and debugging, aiming to improve code quality and expedite the development cycle. 

\subsection{Enhancing Coding Efficiency and Its Impact on Software Development Landscape}
\textbf{1. Leapfrog improvement in programming efficiency:} Our back-of-the-envelope calculation offers a stark contrast between traditional coding productivity and the potential of GenAI-assisted coding. Citing Ray Farias, a Full-Stack Software Engineer at Google, a productive engineer at Google might produce about 100-150 lines of code daily (Farias, 2018)\cite{farias2018linesofcode}, translating to 12-20 lines per hour. In contrast, our research's prompting methodology facilitated the creation of a functional Snake game module—around 100 lines—in merely 5-10 minutes, including debugging and adjustments, which translates to an astonishing 600-1200 lines per hour. This suggests a potential 30 to 100-fold increase in code generation efficiency, offering a glimpse into GenAI's transformative potential for programming practices, despite the limitations of this comparison in fully capturing the nuances of programming efficiency such as code quality and clarity.

\textbf{2. Disproportionate benefits for beginners: }The surge in coding efficiency is especially beneficial for beginners, sidestepping the steep learning curve associated with mastering basic syntax and foundational knowledge. GenAI not only shortens the learning curve but also significantly reduces the time spent on common beginner errors. This shift enables entry-level programmers to focus more on creative and high-value problem-solving tasks, including setting ambitious goals and collaborating with GenAI on new algorithm development, thus expanding the horizons of software development.

\textbf{3. Paradigm shift in programming:} The prompting methodology and GenAI-assisted coding practices introduced by this research indicate a substantial leap in programming productivity, particularly for non-professional programmers. This transition from traditional coding paradigms to a GenAI-assisted approach marks a foundational change in software development, which is also substantiated by T. Eloundou et al. (2023)\cite{eloundou2023gpts}. In the new paradigm, developers adopt a supervisory role, guiding the AI, defining high-level objectives, interpreting suggestions, and ensuring the code meets technical and business goals. The critical competency in this era is evaluating the AI's output for accuracy, efficiency, and security.

\textbf{In summary:} Incorporating GenAI into programming workflows promises to significantly reduce the time-to-market for software products, foster innovation by lowering the entry barriers to software development, and allow non-professional programmers to focus more on complex problem-solving. This shift not only democratizes programming skills but also accelerates the pace of innovation in the software industry.

\subsection{Real-World Impact of GenAI Coding Methodology}
In an effort to extend the reach and applicability of our research, we organized a GenAI Coding Workshop in February 2024. This event served as a practical demonstration of our research-derived prompting methodology, enabling students from varied backgrounds to rapidly acquire and proficiently utilize GenAI tools for programming tasks. The workshop, a collaboration between DeepBlue Technology (Shanghai) and the Global Discovery \& Innovation Academy, catered to college students in their first and second years, including a select group of high school seniors, from diverse academic fields (Appendix 2). Among the 19 participants, most were programming novices with no prior experience in Python and GenAI-assisted coding.

This full-day workshop dedicated two hours to hands-on GenAI programming projects. It kicked off with a 30-minute introduction to GenAI programming principles, followed by a 90-minute of collaborative project work. Attendees swiftly adopted the prompting methodology, venturing beyond Snake to craft classic arcade games like Breakout, Five-in-a-Row, and Tetris. Participants enthusiastically presented their projects during the session wrap-up (Figure 4). Our observations suggested that  after this hour-long GenAI crunch session, even novice GenAI users and beginner-level programmers could, within approximately five minutes, develop a simple yet functional Snake game in Python, comprising typically around 100-200 lines of code, including debugging and final adjustments.
\begin{figure}
    \centering
    \includegraphics[width=1\linewidth]{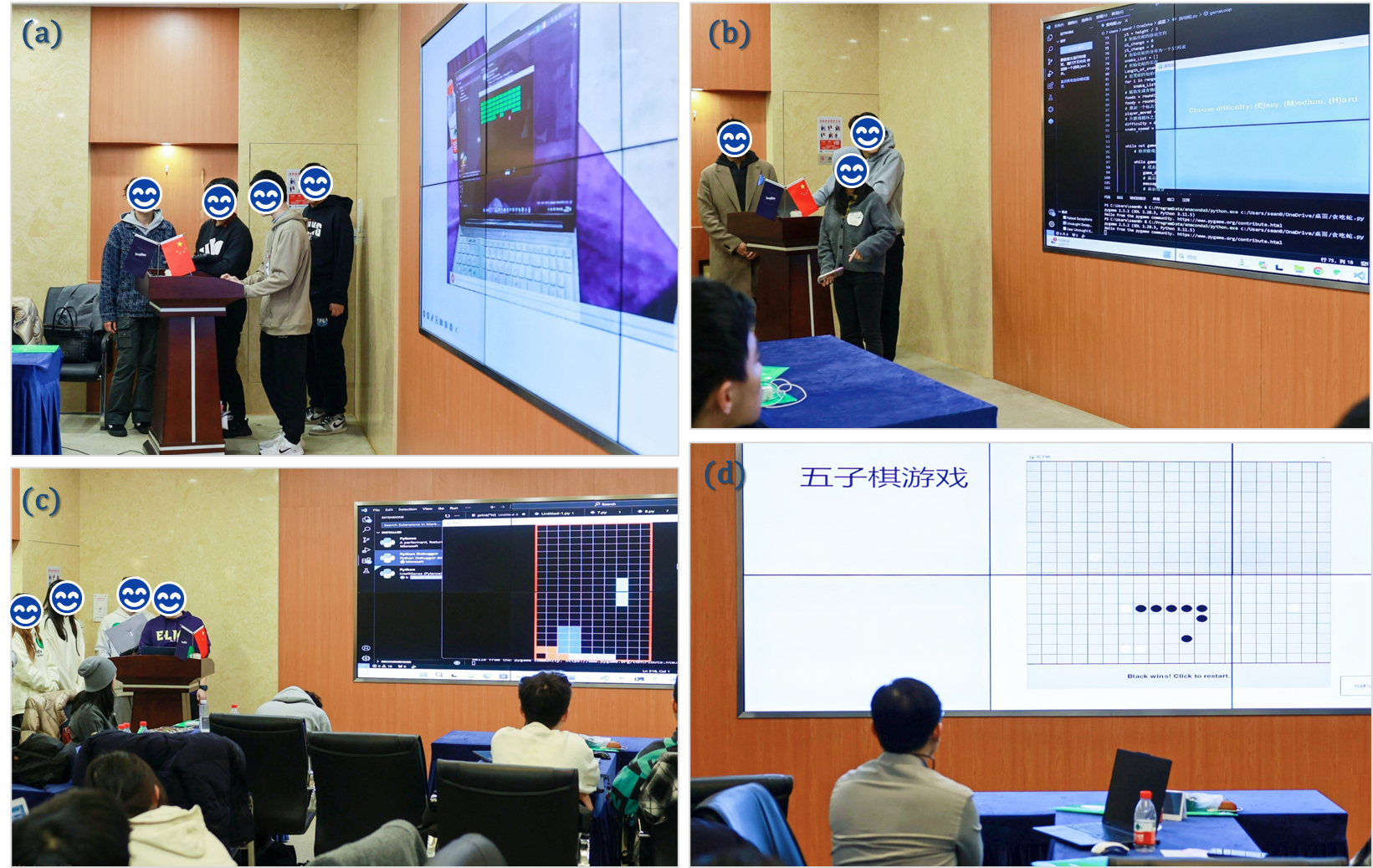}
    \caption{GenAI Coding Workshop at DeepBlue Technology (February 2024) - participant groups present their projects: (a) Breakout game, (b) Snake game with difficulty level selection, (c) Tetris game, and (d) Five-in-a-row game.}
    \label{Figure 4}
\end{figure}

The success of this workshop underscores the adaptability and accessibility of the prompting methodology we've developed, proving that users with minimal prior knowledge can achieve proficiency in under an hour of instruction. This significantly reduces the barrier to entry for both programming and GenAI applications, democratizing skills that were once considered advanced and exclusive. The social implications of this are hence very profound, demonstrating the potential for inclusivity and equality through GenAI applications across diverse demographics. 

\section{Conclusions}
Our study reveals that simple, direct prompts like "generate code for a Snake game in Python" yield the highest one-shot success rate of 90\% for both GPT-4 and GLM-4, with GPT-4 slightly outperforming GLM-4, though the difference is negligible for most users. Furthermore, a preliminary "confirmation round" with a basic question such as "Do you know the classic arcade game Snake?" notably boosts one-shot success rates, likely due to the Chain of Thought mechanism.

Our analysis also indicates a striking 30 to 100-fold increase in code generation efficiency with GenAI, despite some methodological constraints. This highlights GenAI's capacity to significantly boost productivity, especially for beginners by smoothing out the learning curve and cutting down on common mistakes. Consequently, developers transitioning to a supervisory role—guiding AI, setting objectives, and ensuring code alignment with technical and business requirements—signify a paradigm shift in the programming landscape.

The success of our GenAI Coding Workshop further validates that GenAI coding, combined with proper prompting strategy, significantly lowers the barrier to entry for programming and GenAI applications, making previously advanced skills widely accessible. The profound social impact of this development showcases the potential for inclusivity and equality through GenAI applications across a diverse range of demographics.


\bibliography{GenAI_Coding-refs}
\nocite{*}

\begin{acknowledgments}
This research benefited immensely from the support and resources provided by DeepBlue Technology (Shanghai), whose contributions were instrumental in facilitating our research efforts. Special thanks to Ms. PAN Xiaolan and the Global Discovery \& Innovation Academy for their leading role in organizing the GenAI Coding Workshop, a pivotal event that enriched this study. We are also grateful to Prof. BIAN Kaigui and his research team at Peking University for their insightful discussions and invaluable support throughout the research process and in the publication of our findings.
\end{acknowledgments}

\newpage
\section*{Generative AI Assistance Statement}
1. GPT-4, GPTS-Consensus, GPTS-Eddy Master, and GLM-4 were used for writing, editing, coding, brainstorming research ideas, and formatting assistance in this project.

2. About GPTS-Consensus: Your AI Research Assistant. Search 200M academic papers from Consensus, get science-based answers, and draft content with accurate citations. Created by consensus.app.

3. About GPTS-Eddy Master: Expert in editing and translating English/Chinese texts, with a focus on clarity and impact; created by Dr. LI Zehan.

\section*{Appendix}
\appendix
1. representative Python codes from four different levels: 
\href{https://github.com/VonAugustus/SnakeData}{https://github.com/VonAugustus/SnakeData}

2. DeepBlue GenAI Coding Workshop attendees background
\begin{figure}[h!]
    \centering
    \includegraphics[width=1\linewidth]{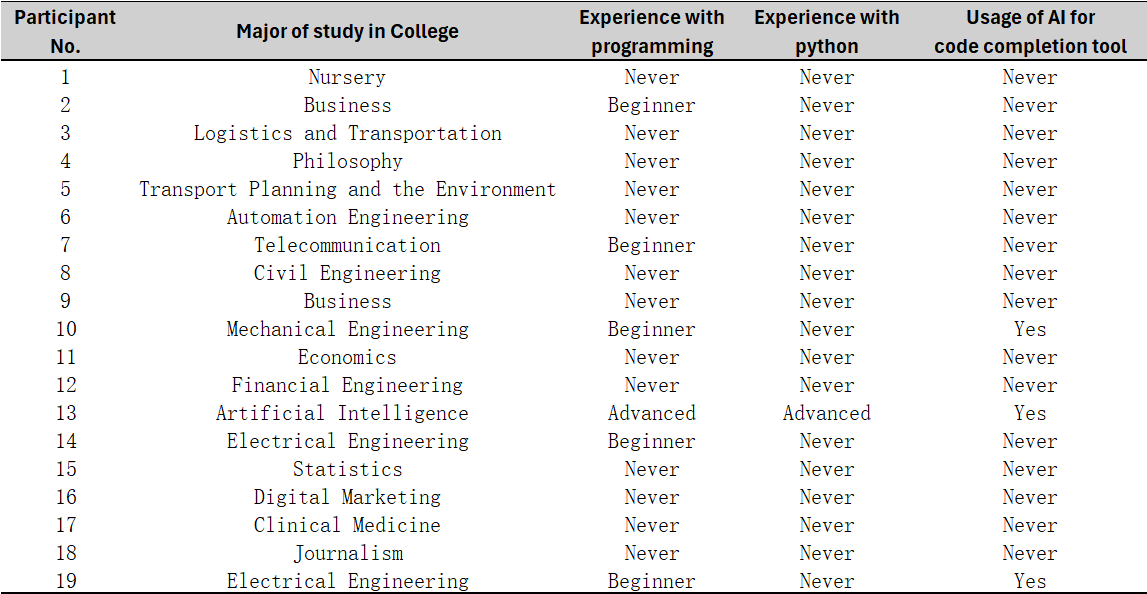}
      \label{Appendix 2}
\end{figure}

\end{document}